\RequirePackage{fix-cm}
\documentclass[smallextended]{svjour3}       
\smartqed  
\usepackage{graphicx}
\usepackage{mathptmx}      
\usepackage{amssymb}
\begin{document}

\title{Speed of the CERN Neutrinos released on 22.9.2011
}
\subtitle{Was stated superluminality due to neglecting General Relativity?}


\author{Wolfgang Kundt
}


\institute{W. Kundt \at
              Argelander-Institut f\"ur Astronomie, Auf dem H\"ugel 71, 53121 Bonn, Germany \\
              Tel.: +49-228-731771\\
              Fax: +49-228-733672\\
              \email{wkundt@astro.uni-bonn.de}           
}

\date{Received: date / Accepted: date}

\maketitle

\begin{abstract}
During the years 2009 to 2011, neutrino beams were fired repeatedly from CERN
towards a detector in Italy's Gran Sasso tunnel, some 4 deg south and 7 deg
east of CERN, at a distance of 730.5 km, in the shape of short bunches of
particles. Their time of flight (2.5 msec) was measured at high accuracy
(nsec) with caesium clocks \cite{Reich2011}. Remarkably, the CNGS team found a
deficit of 61 nsec compared with propagation at the speed of light, and
concluded at superluminal speeds, of order 10$^{-4.6}$. In this communication,
I will argue that this is the first experiment to test Einstein's theory for
the (weak) gravity field of Earth, with the result that the neutrinos
propagated (just) luminally.
\end{abstract}

\section{Speed Measurements in Flat and Curved Spacetime}
\label{ch1}
Newtonian physics loses its applicability when speeds approach the speed of
light, like for neutrinos; we should pass at least to Special Relativity (SR).
In SR, the world lines of CERN and of the Gran Sasso neutrino detector form
two almost straight timelike lines of spatial separation $0.7 \cdot 10^{3}$km, with
a relative velocity \textbf{v} given by corotation with our home planet Earth
at the respective latitudes \{$46^{\circ}$, $42^{\circ}$\} and longitudes \{$6^{\circ}$, $13^{\circ}$\}. The magnitude of v is some 12\% of the corotation speed with Earth
at CERN's (northern) latitude, seen redshifted when viewed from CERN.
Consequently, the Gran Sasso clock moves redshifted w.r.t. the CERN clock,
hence runs more slowly than the CERN clock by a redshift of z = v/c
$\approx 10^{-6.5}$. If SR were the proper spacetime geometry for this
experiment, the neutrino travel time would thus be found shortened by some
$10^{-6.5}$.

But as is well-known, gravity at the surface of Earth modifies clock rates
more strongly than do typical kinematic redshifts (given by \ $\beta$\ :=
v/c). The general-relativistic (GR) counterpart to $\beta^{2}$\ in
\textit{Robertson}'s line element \ --\ \ which describes a generalised
\textit{Schwarzschild} geometry \ --\ is 2m/R: \ \ \
\begin{equation}
\left(\frac{\mathrm{d}\tau}{\mathrm{dt}}\right)^{2}\ \,=\ \,\mathrm{1}-\frac{\mathrm{2m}}{\mathrm{r}%
}-\left(\frac{\mathrm{d\textbf{x}}}{\mathrm{cdt}}\right)^{2}-2\,%
\frac{\mathrm{d\textbf{x}}\cdot\mathrm{\textbf{J}}}{\mathrm{cdt}}
\end{equation}
in application to Earth in which the three PPN parameters $\beta,\gamma,$ and
$\delta$ have been set equal to unity, $\tau$ is proper time (measured, e.g.,
by caesium clocks), m := GM/c$^{2}$ = $10^{0.11}$cm \ is the gravitational
length of Earth, corresponding to its mass $\mathrm{M}=10^{27.78}$g, r := radial
coordinate, and \ \textbf{J} := $-4\int\frac{\mathrm{d\textbf{x}}}{\mathrm{cdt}%
}\frac{\mathrm{dm}}{\mathrm{r}}$ is the vector potential for the spin motion of
Earth; \textbf{J} is understood as the retarded volume integral over Earth; it
describes the \textit{Lense-Thirring effect}. In our application, m/r is
$\gtrsim$10$^{-9}$ (for r $\approx$ R), and $\beta$ := dx/cdt $\gtrsim$
$10^{-6}$ holds for our comoving clocks, so that J is negligibly small
($\approx 10^{-15}$), and the line element simplifies to: \ (d$\tau
$/dt)$^{2}$ = (1 - 2m/r - $\beta^{2}$), \cite{KrotscheckKundt1983}\cite{Will1993}.

We now see that for the curved GR geometry of Earth described approximately by
the truncated Robertson line element, all we have to do is replace the
kinematic potential $\beta^{2}$ of the SR approach by the full gravitational
potential \ 2m/r + $\beta^{2}$ ($\gtrsim$ 2m/r), in order to get a fair
estimate of the deviations of true (geometric) velocities from coordinate
velocities. Even so, this task is far from trivial because near a rotating
body, there do not exist global spacelike hypersurfaces (of fixed time)
\ --\ \ just remember G\"{o}del's cosmological model \ --\ \ nor are photon
paths described by straight lines (as in SR). We would be forced into lengthy
calculations applied to a timelike triangle spanned by a neutrino's null
geodesic, by its (curved) 3-space projection, and by the (curved, timelike)
worldline of the detector: Are the observed 61 nsec deviation a geometric
effect? According to \cite{KrotscheckKundt1978}, the answer is a clear
''no'': the basic equations of motion (for elementary particles) do not allow
for tachyonic solutions.

But we can argue more directly: We can compare our GR problem with the
corresponding SR problem considered above, in which we replace the small
kinematic redshift $\beta$\ ($\gtrsim 10^{-6}$) by the much larger
gravitational redshift $\sqrt{\mathrm{2m}/\mathrm{R}}\approx 10^{-4.4}$ in
Robertson's line element, a shift which is just slightly larger than the
relative time deficit ($10^{-4.6}$) measured recently by the two clocks. This
estimate leads me to the conclusion that the caesium clocks have measured a
coordinate effect, not an excess of the neutrinos' speed\ over luminal. Note
that one nsec in a day means a clock uncertainty of smallness 10$^{-14}$! General
Relativity must no longer be ignored in terrestrial high-time-resolution measurements.

\section{Summary}
\label{summary}
When the terrestrial Loran (Long-range-navigation) system was improved from a
timing accuracy of $\mu$sec to nsec, we entered the era of
general-relativistic kinematics on Earth.
This experiment is the first to have shown it.

\begin{acknowledgements}
I am indebted to Ole Marggraf for support, encouragement, and help with the electronics.
\end{acknowledgements}



\end{document}